# Biodynamic Analysis of Alpine Skiing with a Skier-Ski-Snow Interaction Model

Nan Gao[1], Huitong Jin[1], Jianqiao Guo[2, *], Gexue Ren[1] and Chun Yang[1, *]

[1] *Department of Engineering Mechanics, Tsinghua University, Beijing, China.*
[2] *MOE Key Laboratory of Dynamics and Control of Flight Vehicle, School of Aerospace Engineering, Beijing Institute of Technology, Beijing, China.*

**Abstract**

This study establishes a skier-ski-snow interaction (SSSI) model that integrates a 3D full-body musculoskeletal model, a flexible ski model, a ski-snow contact model, and an air resistance model. An experimental method is developed to collect kinematic and kinetic data using IMUs, GPS, and plantar pressure measurement insoles, which are cost-effective and capable of capturing motion in large-scale field conditions. The ski-snow interaction parameters are optimized for dynamic alignment with snow conditions and individual turning techniques. Forward-inverse dynamics simulation is performed using only the skier's posture as model input and leaving the translational degrees of freedom (DOFs) between the pelvis and the ground unconstrained. The effectiveness of our model is further verified by comparing the simulated results with the collected GPS and plantar pressure data. The correlation coefficient between the simulated ski-snow contact force and the measured plantar pressure data is 0.964, and the error between the predicted motion trajectory and GPS data is 0.7%. By extracting kinematic and kinetic parameters from skiers of different skill levels, quantitative performance analysis helps quantify ski training. The SSSI model with the parameter optimization algorithm of the ski-snow interaction allows for the description of skiing characteristics across varied snow conditions and different turning techniques, such as carving and skidding. Our research advances the understanding of alpine skiing dynamics, informing the development of training programs and facility designs to enhance athlete performance and safety.

## 1. Introduction

Alpine skiing, an exhilarating sport characterized by high speeds and complex movements, demands a profound understanding of biomechanical principles to optimize performance and minimize injury risks. Biomechanical analysis is, therefore, crucial in both the training of alpine skiers and the design of skiing facilities.[1] Studies have highlighted the importance of musculoskeletal modeling and simulations in calculating joint torques and muscle forces that are otherwise impossible to measure directly.[2] Existing research has utilized motion capture system, GPS, electromyography, and ergometer to collect experimental data, and musculoskeletal models have been constructed for simulations across various types of skiing, including cross-country skiing, ski jumping, seated double poling, and alpine skiing.[3–6] The mechanics of ski-snow interaction form the foundation for understanding skier actions and must be carefully considered in modeling efforts.[7] Thus, kinematic data of the skier and external force data are pivotal model inputs in musculoskeletal analysis. To acquire external force data, Holmberg and Lund conducted pole force measurements to perform full-body musculoskeletal simulations based on motion capture.[8] Lee *et al.* used wearable inertial sensors and plantar pressure insoles to acquire kinematic data and normal compression forces on foot, respectively, as fundamental inputs for analyzing joint forces and moments in ski carving turns.[9] Notably, however, plantar pressure insoles are limited in detecting the one-dimensional force applied to the bottom of the ski boots, ignoring the force exerted by the boot cuff. Hirose *et al.* calculated joint torques in ski turns using inertial and 6-axis force sensor data.[10] Kurpiers *et al.* employed high-speed cameras and 6 degrees of freedom (DOFs) force sensors to collect data on freestyle mogul skiing.[11] Despite the value of 3D dynamometers, their use can increase the skier's standing height and add weight, potentially impacting performance.[12]

Modeling ski-snow contact behavior offers an alternative to using external force sensors, enhancing comfort and convenience during testing. Chen and Qi used a spring-damper element to describe penetration force and classical Coulomb friction for friction.[13] Heinrich *et al.* developed a musculoskeletal model enabling only 2D simulation in the sagittal plane.[14] Gong *et al.* established a mannequin model with skis in AnyBody software to analyze landing mechanism.[15] Heinrich *et al.* considered penetration, shear, and friction forces between the ski and snow under a constant friction coefficient.[16] However, these current ski-snow contact models do not account for variability in snow conditions and turning techniques, such as carving or skidding, and they lack validation with experimental data or are limited in their ability to simulate real-world skiing scenarios.

To address these gaps, our study aims to develop a skier-ski-snow interaction (SSSI) model that integrates a 3D full-body musculoskeletal model, a ski model, a ski-snow contact model, and an air resistance model. The objectives of this study are to: 1, Develop an experimental method to collect kinematic and kinetic data of alpine skiing that is convenient, cost-effective, and capable of capturing motion in large-scale and extreme field conditions; 2, Establish a ski-snow interaction model with parameters personalized based on experimental kinematic data to achieve dynamic alignment with snow conditions and the skier's turning technique; 3, Develop a SSSI model based on the aforementioned ski-snow interaction model to accurately reproduce the mechanical processes of skiing; 4, Identify kinematic and kinetic parameters for the quantitative analysis of ski turning techniques. This comprehensive approach aims to bridge the gaps in current biomechanical models, providing a deeper understanding of alpine skiing mechanics and informing the development of training programs and facility designs to enhance athlete performance and safety.

## 2. Methods
### 2.1 Experiments
Two healthy skiers were recruited for the experiment, including one beginner (age: 47 years, height: 1.65 m, weight: 70 kg, Association of Italian Ski Instructors (AMSI) certification for Level Ⅲ) and one intermediate (age: 34 years, height: 1.73 m, weight: 84 kg, AMSI certification for Level Ⅴ). After warming up, participants donned a wearable skiing motion capture system (as shown in Fig.1), comprising an inertial motion capture system equipped with 17 inertial sensors (the Noitom Perception Neuron, Noitom Technology Ltd., Beijing, China), a high-precision global positioning system (Pike, ReliableSense, Jiangsu, China), and a backpack containing a data-collecting laptop. The inertial sensors were strapped securely to the participants' limbs, and the GPS device was positioned in the backpack with its antenna exposed to ensure signal quality. Prior to testing, participants underwent static calibration of the inertial system. They were then instructed to perform linked turns at their self-selected speed, emphasizing their highest skiing skills. The tests were conducted on a compacted, freshly groomed ski run with an average inclination angle of 16° at Wanlong Ski Resort, Hebei, China.

For each participant, full-body motion data and GPS data were collected synchronously, with the IMU sampling at 60Hz and the GPS at 10Hz. For model validation, a pair of plantar pressure measurement insoles (Beijing Speed Smart Technologies Co., Ltd., Beijing, China), each containing 16 pressure sensors with a measurable range of 0-5MPa, was incorporated into the beginner's wearable skiing motion capture system. These pressure insoles were inserted into the ski boots and synchronously recorded foot pressure data at a sampling frequency of 100Hz. After validating the model using data of the beginner, the insoles became redundant as ski-snow contact forces could be correctly simulated. Consequently, during the test of the intermediate, only the IMU and GPS were utilized for enhanced comfort and convenience.

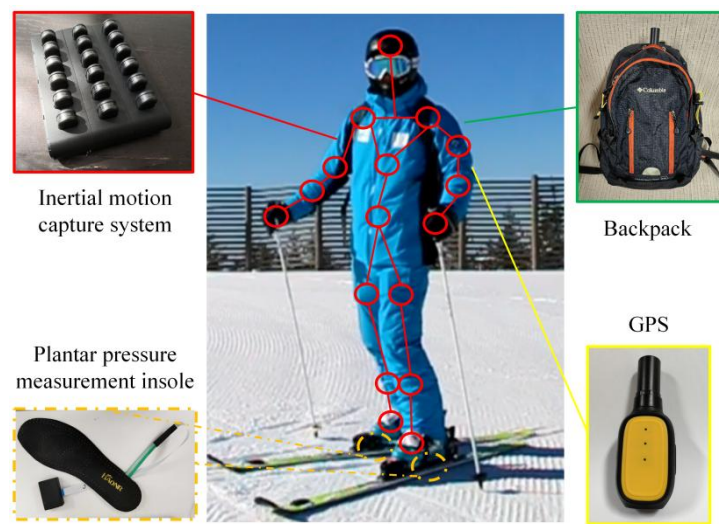

**Fig 1** The wearable skiing motion capture system, including an inertial motion capture system, a high-precision GPS, a backpack containing a laptop, and a pair of plantar pressure measurement insoles, which were used exclusively for model validation.

**2.2 Establishment of the SSSI model**

The SSSI model (Fig.2a) consists of 4 parts: the full-body musculoskeletal model, the ski model, the ski-snow contact model and the air resistance model.

**2.2.1 The musculoskeletal model**

The musculoskeletal model was developed based on the OpenSim full-body lumbar spine (FBLS) model,[17] comprising 21 rigid segments. The model has a total of 31 DOFs: 6 describe the position and orientation of the pelvis relative to a fixed frame of reference; 3 at each hip and shoulder joints; 2 (elbow flexion and forearm rotation) at each elbow joint; 1 at each of the knee, ankle and wrist joints; and 3 at the lumbar joint between the torso and pelvis. The wrist deviation joint was fixed. We excluded the mass of the pole and neglected the impact of pole plant. Additionally, the subtalar and metatarsophalangeal joints in each lower extremity were immobilized due to ski boots, allowing solely plantarflexion and dorsiflexion at the ankle joint.

The skier's motion was actuated by 284 muscles (23 per leg and 238 actuating the lumbar joint). The major muscle groups include: gluteus maximus, psoas major, gluteus medius, hip abductors, quadriceps femoris, hamstrings, triceps surae, and tibialis anterior in lower limbs; erector spinae, rectus abdominis, external obliques, internal obliques, multifidus, quadratus lumborum, psoas major, and latissimus dorsi in lumbar spine.

**2.2.2 The ski model**

The ski model (Fig.2b) was established as a Timoshenko beam based on geometrically exact beam formulation (GEBF). Each ski was discretized into 50 flexible beam elements with varying width and thickness. Each beam element is described by generalized coordinates, consisting of the position vector and the orientation.[18] The nodes were assigned 6 DOFs, with the intermediate node of the ski assumed to be fixed to the foot. The ski's geometry was determined through three-dimensional scanning tests conducted on a recreational racing ski (Atomic Redster Doubledeck 3.0 SL). The density was calculated based on the total mass and geometry of the ski. Given the assumption of material homogeneity and isotropy, we assessed the flexural stiffness (EI) and vibration properties of the ski through 3-point bending tests and free vibration tests. Torsional stiffness (GJ) was omitted from measurement due to its negligible influence on computational results.[19] The Poisson's ratio was set to 0.3, consistent with that of wood, the predominant material used in skis.[20]

**2.2.3 The ski-snow contact model**

The snow surface was assumed to be compacted, homogeneous and isotropic. Its geometry was modeled as an ideal slope with a constant inclination angle, determined by the average slope inclination angle measured by GPS. The possible contact geometry between the ski edge and the snow surface was simplified to multiple spheres, arranged side by side along the ski. Each ski had 50 contact detection points on each edge, resulting in a total of 200 contact detection points. A dissipative contact model[21] (Fig.2c) was utilized to determine the ski-snow penetration force, which acts perpendicular to the snow surface.

$$F_N = k\delta^{m_1} + c\frac{\dot{\delta}}{|\dot{\delta}|}|\dot{\delta}|^{m_2}\delta^{m_3}$$

where $k$ and $c$ represent the spring and damping coefficients, respectively. $\dot{\delta}$ denotes the time differentiation of the penetration depth $\delta$. The exponents $m_1$ and $m_2$ generate a non-linear contact force, while the exponent $m_3$ produces an indentation damping effect.

The force within the plane of the snow surface can be resolved into two components[22]: friction, opposing the direction of travel, and the resistance force from snow compression, acting parallel to the snow surface and perpendicular to the direction of travel, facilitating turning. The anisotropy of friction between the ski and the snow surface can be characterized as:

$$\begin{cases} f_t = \mu_t \sum_{i=1}^{n} F_N^i \\ f_n = \mu_n \sum_{i=1}^{n} F_N^i \end{cases}$$

where $f_t$ and $f_n$ are the resultant forces of tangential and normal friction, respectively, acting on calcaneus. The friction coefficients $\mu_t$ and $\mu_n$, strongly influenced by the properties of the snow[23][24], were meticulously optimized for each test scenario. This optimization process was characterized by refining the parameters until convergence criteria were met, indicated by the simulated center of mass trajectory closely aligning with the experimental data.

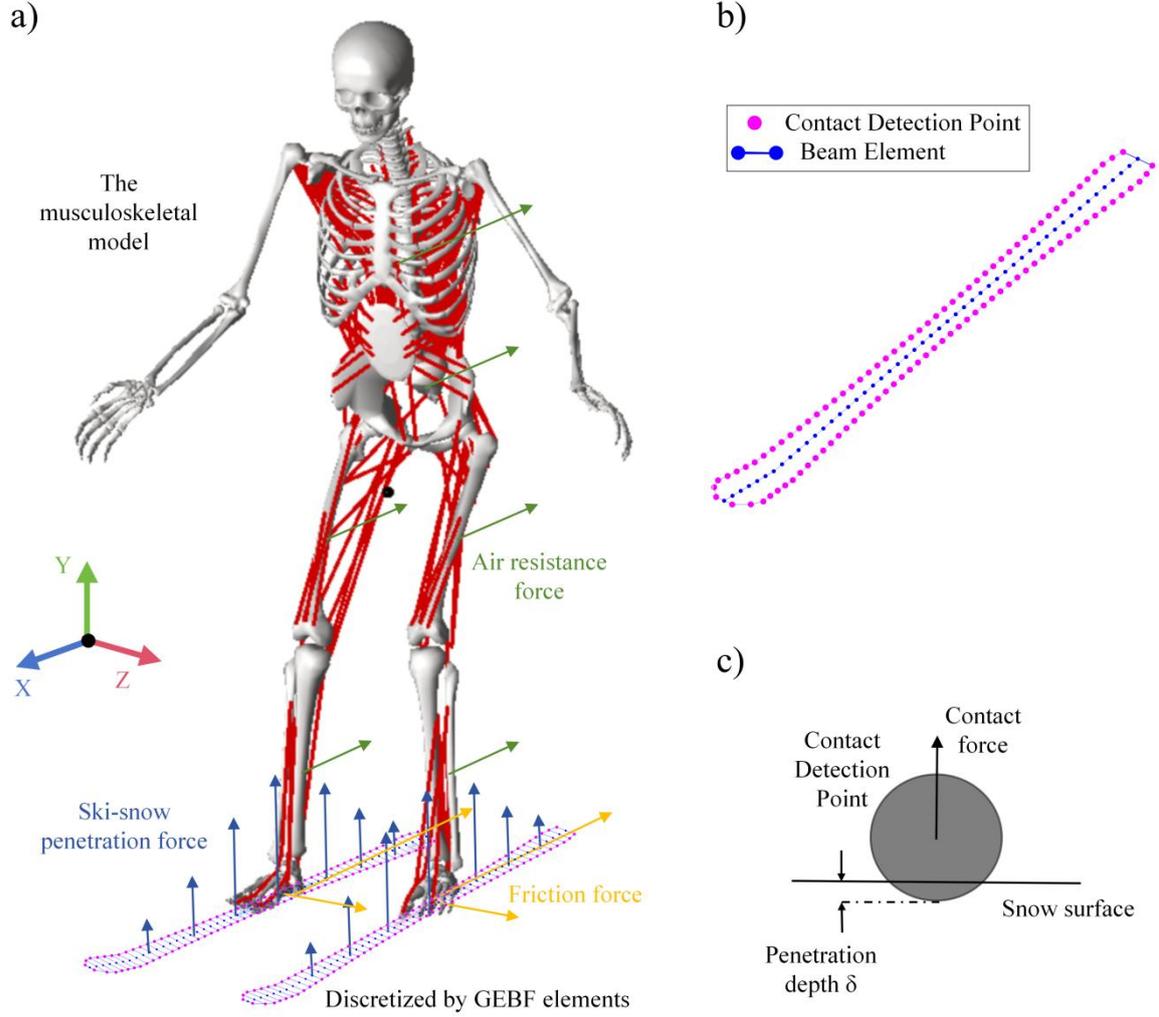

**Fig 2** a) The SSSI model. The model comprised the musculoskeletal model, the ski model, the ski-snow contact model and the air resistance model. b) The ski model. c) The dissipative contact model.

### 2.2.4 The air resistance model

The model consists of 14 anatomical segments,[25] including the pelvis, torso, right and left thighs, legs, feet, arms, forearms, and hands. Air resistance acting on these segments was accounted for using the following formula:[24]

$$F_{Di} = \frac{1}{2} C_D \rho A_i v^2, \quad i = 1, 2, \ldots, 14$$

In this equation, $C_D$ represents the drag coefficient, $\rho$ denotes the density of air, $A$ signifies the frontal area of the body exposed to airflow, and $v$ is the velocity.

### 2.3 Simulation framework

Forward-inverse dynamic analysis was employed to track the experimental data and simulate the joint moments of the skier. To maintain dynamic consistency, only the skier's posture was used as input, and the translational DOFs between the pelvis and the ground were left unconstrained. The joint angles obtained from the inertial motion capture system, denoted as $\theta(t)$, served as motion constraints:[26]

$$C_{mot} = \theta(q) - \theta(t) = 0$$

The governing equations for the forward-inverse dynamics of the musculoskeletal system can be expressed as follows:

$$\begin{cases} M(q,t)\ddot{q} - Q(q,\dot{q},t) + C_q^T \lambda = 0 \\ C(q,t) = [C_{joint}^T, C_{mot}^T]^T = 0 \end{cases}$$

where $M$ is the mass matrix of the system, $q$ is the vector of the generalized coordinates, $Q$ is the generalized force vector, and $\lambda$ is the Lagrange multipliers. The vector $C$ includes constraints of the entire system, which consist of anatomical joints $C_{joint}$ and exerted motion constraints $C_{mot}$. Then, the vector of the joint torques $M_{ID}$ can be calculated based on the following expression:

$$M_{ID} = \left(\frac{\partial \theta}{\partial q}\right)^{-T} \left(\frac{\partial C_{mot}}{\partial q}\right)^T \lambda_{mot}$$

**2.4 Data analysis**

Following data acquisition, the friction parameters were optimized by performing forward-inverse dynamics analysis using the acquired joint angle data, with iterative adjustments to the friction parameters at each step. Once the convergence criteria were satisfied, the optimal friction parameters were determined. The model's validity was confirmed by comparing the simulated center of mass trajectory and ski-snow contact forces with corresponding experimental results. Subsequently, the quantitative analysis of two skiers was conducted.

**Raw data acquisition**

The raw data obtained from the inertial motion capture system were converted into Biovision Hierarchy (BVH) format using Axis Studio Software (Noitom Technology Ltd., Beijing, China). This format delineates the hierarchical structure of the skeleton and encompasses raw joint angle data over time. The raw foot pressure data were obtained by summing the data from all the pressure sensors on each insole. Subsequently, both the raw joint angle data and the raw foot pressure data were filtered using a 4th-order Butterworth filter with a 6 Hz low-pass cutoff frequency.

**Friction parameters optimization**

Alpine skiing relies fundamentally on gravity, snow-ski friction, and air-body friction, all of which are influenced significantly by the skier's posture. When simulating alpine skiing, the friction between snow and skis is critical and varies with temperature, ski direction, and skier's edging skills, showing anisotropic behavior. Consequently, it is imperative to optimize friction parameters rather than treating them as fixed values. The friction parameters were optimized using experimental data to identify the most suitable parameters for the given test scenario. This process can be described as a constrained optimization problem:

$$\min f = \int_0^t \|\bm{x}_{\exp} - \bm{x}_{\text{sim}}\|^2 dt$$

$$s.t. \begin{cases} \bm{M}(\bm{q},t)\ddot{\bm{q}} - \bm{Q}(\bm{q},\dot{\bm{q}},t) + \bm{C}_q^T \bm{\lambda} = \bm{0} \\ \bm{C}(\bm{q},t) = [\bm{C}_{joint}^T, \bm{C}_{mot}^T]^T = \bm{0} \\ \mu_{\min} \leq \mu_t \leq \mu_{\max} \\ \mu_{\min} \leq \mu_n \leq \mu_{\max} \end{cases}$$

Where $f$ is the objective function, denoting the error between the experimental and simulated motion trajectory data. The equality constraints are the governing equations for the forward-inverse dynamics of the musculoskeletal system. The bounds on decision variables $\mu_t$ and $\mu_n$ were set between $\mu_{\min} = 0.01$ and $\mu_{\max} = 0.99$, corresponding to the physical limits of friction parameters.

The Sequential Quadratic Programming (SQP) algorithm was employed to address this non-linear optimization challenge. Initial guesses for the tangential and normal friction parameters were established as $\mu_t = 0.1$ and $\mu_n = 0.5$, respectively, and the model was driven by joint angle data, followed by forward-inverse dynamics simulations. The discrepancy between the simulated motion trajectory derived from forward-inverse dynamics and the experimental motion trajectory obtained via GPS was used to iteratively update the friction parameters. This iterative process continued until convergence criteria were met, thereby minimizing trajectory errors and obtain the optimal values of $\mu_t$ and $\mu_n$.

**Model validation**

The model's accuracy was validated by comparing simulated center of mass trajectory and ski-snow contact force with corresponding experimental data. The simulation results were derived from forward-inverse dynamic simulation with the SSSI model. Discrepancies between simulated and GPS trajectories were quantified as percentage trajectory error, defined as the average trajectory error divided by the length of the trajectory. The simulated ski-snow contact force and the experimental foot pressure were normalized to the pressure ratio, defined as the ratio of the left foot pressure to the total foot pressure. Then, the disparity between the simulated ski-snow contact force and the experimental foot pressure was assessed using the correlation coefficient and the root mean squared difference (RMSD) of the pressure ratio.

**Quantitative analysis of turning technique of Alpine ski**

In consequence to the model validation, we selected two linked turns (one right turn and one left turn) from both beginner and intermediate skiers as examples for analysis. A skiing turn is delimited by two consecutive transition points, which occur when the turning radius reaches its maximum. We divided a turn into three equal time-based phases: initiation, control, and completion. For each subject, we optimized the friction parameters and then used the optimal values of $\mu_t$ and $\mu_n$ to conduct inverse dynamics simulations, obtaining kinematic and kinetic parameters of the skiing maneuver.

The musculoskeletal model was implemented using our in-house code developed in C/C++[26].[26] The code for data processing was written in MATLAB R2022b (MathWorks, Inc., Natick, MA). All simulations were conducted on a 64-bit desktop computer with 16 GB RAM operating at 4.65 GHz.

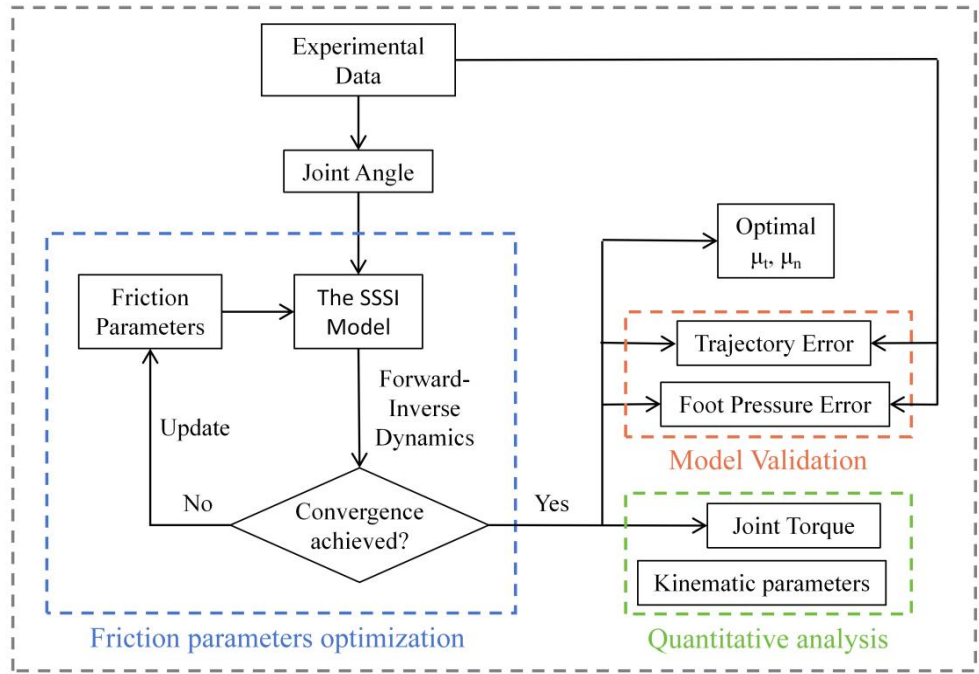

**Fig 3** Optimization and simulation procedure. The friction parameters were optimized through forward-inverse dynamics, with updates applied at each step. Once the convergence criteria were satisfied, the optimal friction parameters were obtained. The model was then validated by comparing trajectory and foot pressure errors between experimental and simulated data. Subsequently, a quantitative analysis of kinematic and kinetic parameters was performed.

## 3. Results

### 3.1 Friction parameters derived from the optimization

The optimal friction parameters ($\mu_t$ and $\mu_n$) and the trajectory error are shown in Table.1. The friction parameter in the tangential direction was 0.06 for the beginner and 0.19 for the intermediate skier. These results align with the experimental friction coefficients reported by Buhl *et al.* [27], and Hasler M *et al.* [28]. The friction coefficient in the tangential direction can be influenced by factors such as temperature, speed, normal load,[23] ski's edge angle and ski properties.[29] The warm weather during the intermediate skier's test, with temperatures slightly above 0 degrees Celsius causing snow melting, may contribute to the much higher friction coefficient observed compared to the beginner.[27]

In the normal direction, the friction parameter was 0.55 for the beginner and 0.9 for the intermediate skier. This difference can be attributed to variations in ski turning techniques: the beginner exhibited skidding, while the intermediate skier employed approximate carving. The higher normal direction friction coefficient of the intermediate suggests that the snow surface provided greater resistance when the ski held on the snow compared to when it slipped.

### 3.2 Validation of the SSSI model

Validation of the forward-inverse dynamics results is shown in Fig.4. The average trajectory error between the simulated and experimental trajectories is 0.134 m, indicating a high degree of consistency between them. The correlation coefficient and RMSD of the foot pressure ratio are 0.964 and 0.162, respectively, demonstrating the accuracy in both phase and amplitude.

**Table 1** The optimal friction parameters and the percentage trajectory error of the beginner and the intermediate.

|  | $\mu_t$ | $\mu_n$ | Percentage Trajectory Error | Piste Inclination Angle/° |
|---|---|---|---|---|
| Beginner | 0.06 | 0.55 | 0.7% | 15.4 |
| Intermediate | 0.19 | 0.9 | 0.6% | 16.5 |

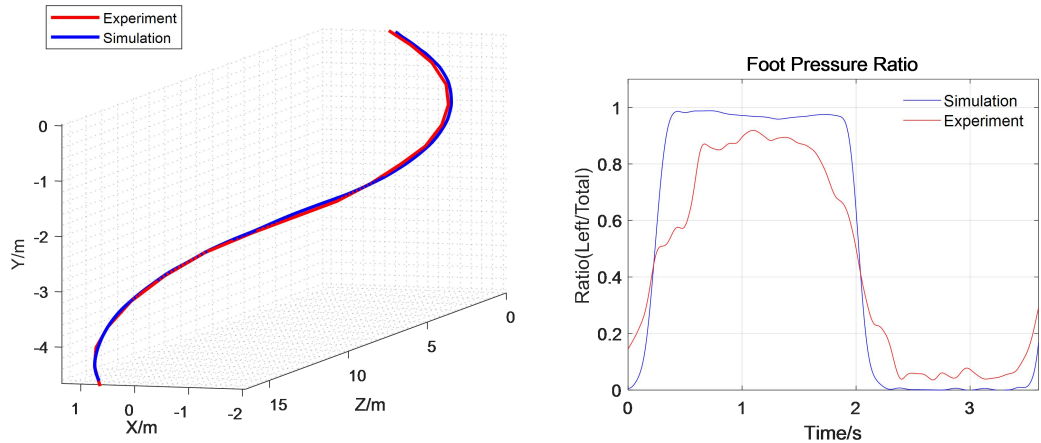

**Fig 4** Comparison of the simulated and experimental skiing trajectory (a) and foot pressure ratio (b) during two linked turns.

### 3.3 Kinematic parameter of the skiers

Kinematic parameters of the skiers were then extracted from the ski process reconstructed by the model. (see Supplement) Figure 5 presents the kinematic data for beginner and intermediate skiers during two linked turns. The duration of the turns was 3.6 s for the beginner and 5.0 s for the intermediate skier. The maximum hip flexion angle of the outside leg was 55° for the beginner and 65° for the intermediate, indicating a greater inclination of the intermediate in the coronal plane. The point of postural shift, indicated by the intersection of left and right hip flexion angles, occurred 0.2 s after the transition time for the beginner, whereas it coincided with the transition time for the intermediate, suggesting optimal timing of postural shift.

### 3.4 Trajectory in the transverse plane

Fig.6 illustrates the center of mass (CoM) trajectory in transverse plane, along with positions of both skis, lines passing through shoulders, and lines passing through hips, all at each instant in time. The average turning radius of the CoM trajectory was 7.7 m for the beginner and 17.6 m for the intermediate. Additionally, the mean speed was 5.5 m/s for the beginner and 11.5 m/s for the intermediate. The average angle between the two skis during the turn is 9° for the beginner and 2° for the intermediate, suggesting the beginner performed wedge turns while the intermediate executed parallel turns. The steering angle of the ski, defined as the angle between the direction of travel and the direction in which the ski is pointed, is significantly larger for the beginner (averaging 25° and peaking at 51°) compared with the intermediate (averaging 6° and peaking at 17°). This discrepancy suggests greater skidding and a more pronounced deceleration effect for the beginner. In the case of the beginner, the CoM consistently resided on the inside of one ski throughout the turn, whereas for the intermediate, the CoM was always positioned inside both skis. This proficiency, referred to as 'crossover' by instructors, marks the transition point to advanced skiing.[22] Lines passing through shoulders and lines passing through hips were used to represent the rotation of pelvis and torso in the transverse plane, aiding in the analysis of one of the most subtle and important facets of body alignment in alpine skiing: counter.

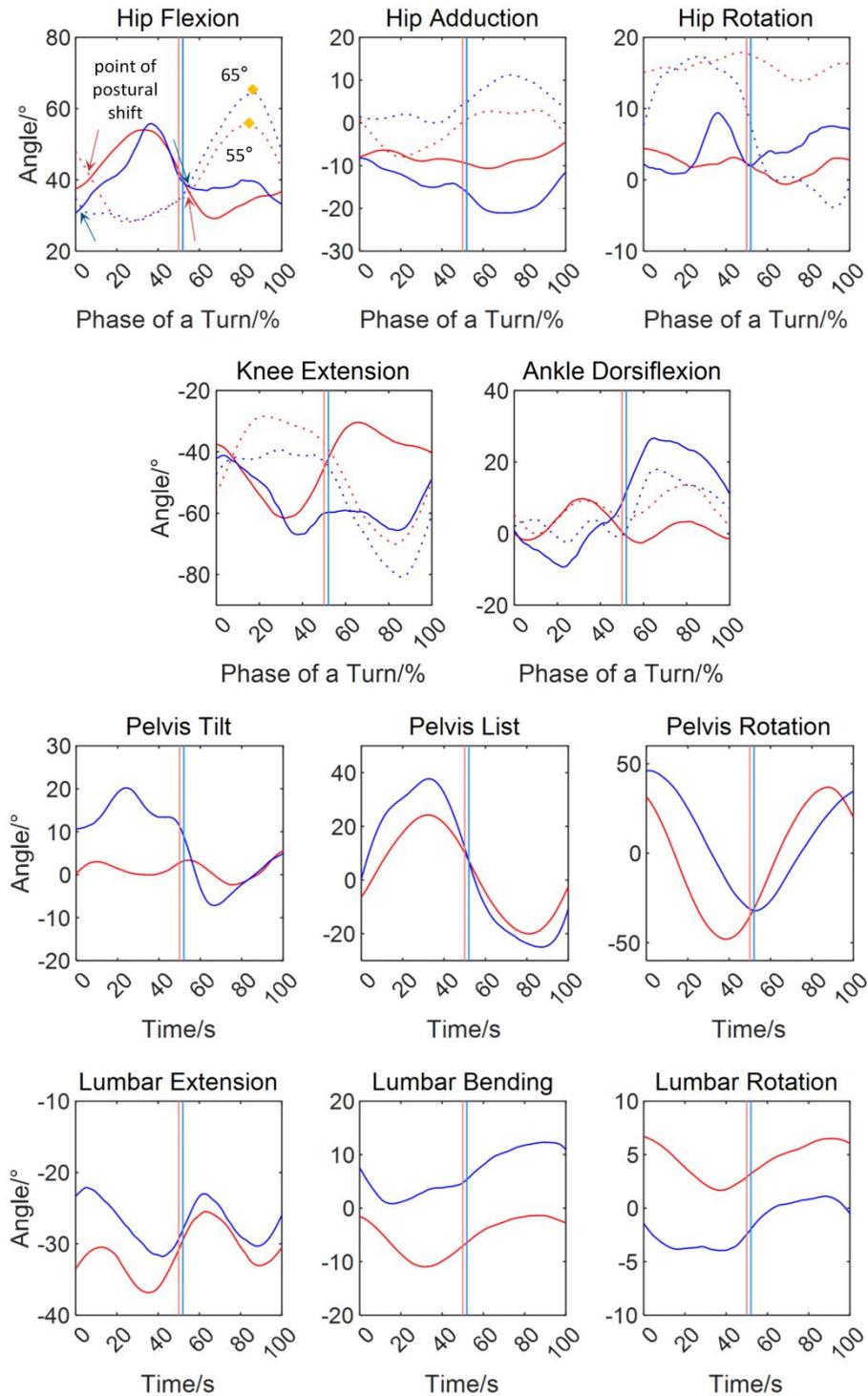

**Fig 5** Kinematic data of the beginner and the intermediate. Red and blue lines indicate data of the beginner and the intermediate, respectively. Solid and dashed lines represent joint angles for the right side, and left side, respectively. Salmon and dodger blue lines indicate transition time between turns of the beginner and the intermediate, respectively. Positive angles denote: hip flexion, hip adduction, hip internal rotation, knee extension, ankle dorsiflexion, pelvis backward tilt, pelvis right list, pelvis left rotation, lumbar extension, lumbar right bending, and lumbar left rotation.

**3.5 Ski-snow contact forces**

The ski-snow contact forces are depicted in Fig.7. The average force ratio on the outside ski, defined as the force on the outside ski divided by the total forces on both skis during the control phase, was 0.99 for the beginner and 0.80 for the intermediate, suggesting that the beginner placed almost all of the weight on the outside ski while the intermediate balanced on both skis. The result of the intermediate is consistent with those reported by Stricker et al. [12], who used a 3D-dynamometer and detected a maximum force of 1800 N on outside ski and 800 N on inside ski, resulting in a ratio of 69%. Similarly, Ogrin et al. [30] observed average force ratios of 65% on outside ski during left turns and 60% during right turns among elite skiers, based on pressure insoles measurement. Additionally, Bon et al. [31] analyzed kinetic parameters of alpine skiing using pressure insoles, reporting a ratio of 78% in parallel turns. These findings suggest that force ratios on the outside ski ranging from 60% to 80% may be considered appropriate for balancing on both inside and outside skis, indicating that the intermediate effectively managed the balance whereas the beginner demonstrated less proficiency in this aspect. The average total forces on both skis in the control phase were 1.19 body weight (BW) for the beginner and 1.40 BW for the intermediate, pointing to a higher centrifugal force experienced by the intermediate. The shift of force between skis, indicated by the intersection of left and right ski-snow contact forces, occurred after the transition time for both the beginner and the intermediate. For the beginner, this shift occurred 0.24 s after transition, constituting 13.3% of the turn duration, whereas for the intermediate, it occurred 0.13 s after transition, representing 5.2% of the turn duration.

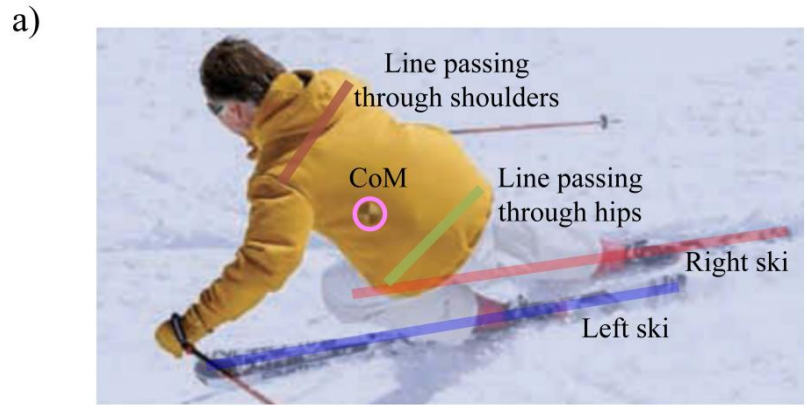
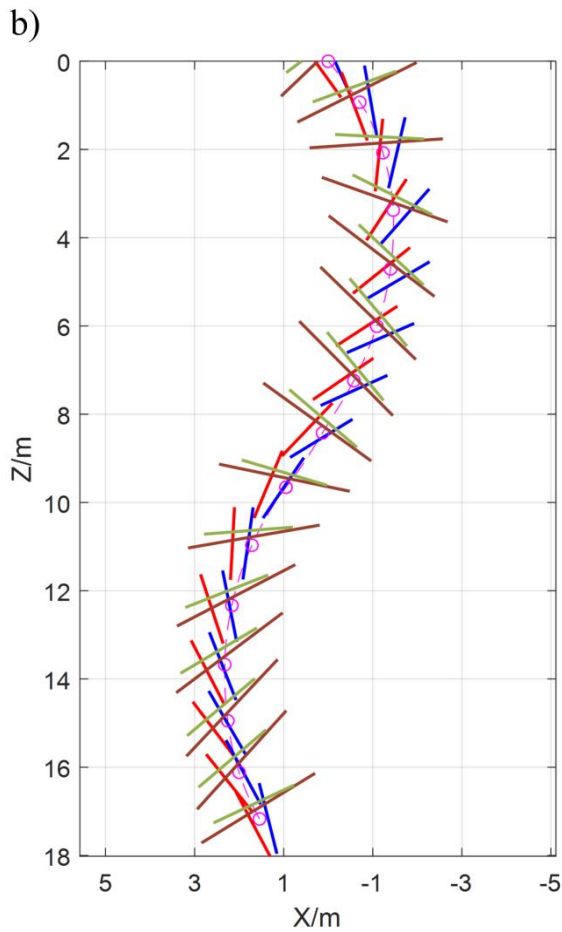
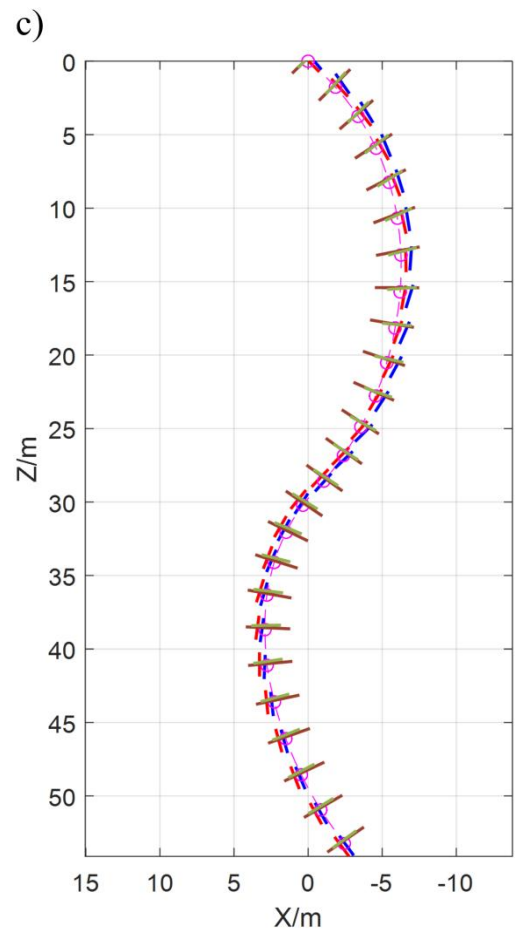

**Fig 6** Trajectories in the transverse plane for the beginner (b) and the intermediate (c). The legend is shown in (a). The slope's fall line is oriented along the positive direction of the Z-axis. The trajectory of the CoM is represented by the dashed purple line, with purple circles denoting its positions at distinct time points. The red and blue solid lines represent the positions of the right and left skis, respectively, at these time points. Additionally, brown and green solid lines depict lines passing through shoulders and lines passing through hips, respectively, at these time points.

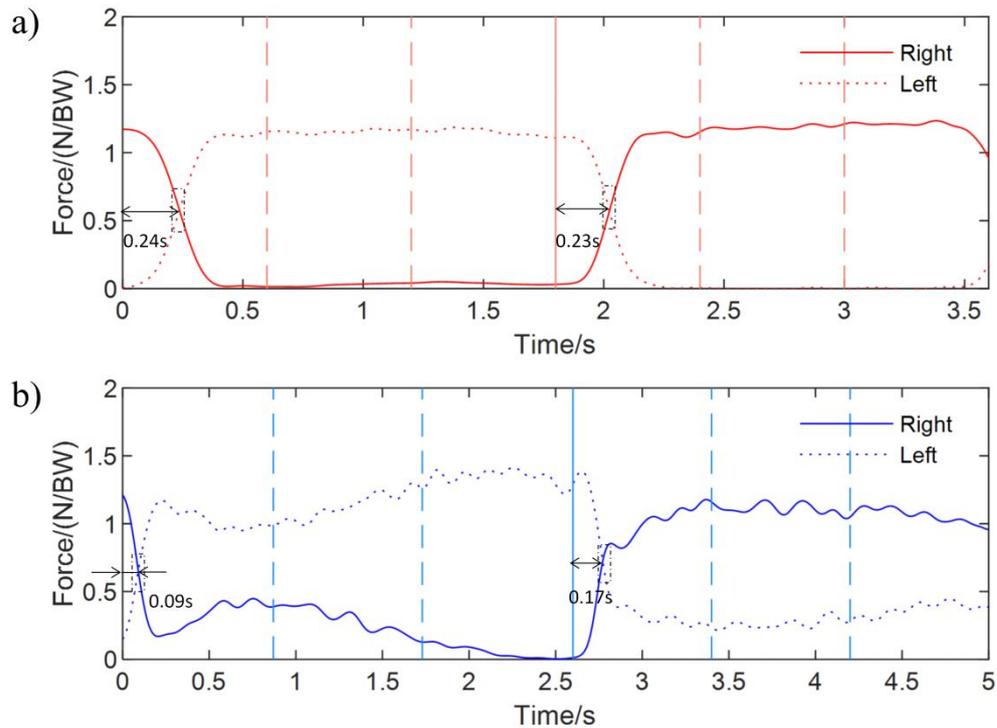

**Fig 7** Ski-snow contact force for the beginner (a) and the intermediate (b), scaled to body weight (BW). Solid and dashed lines represent the forces on the right and left skis, respectively. Solid salmon and dodger blue lines indicate transition time between turns of the beginner and the intermediate, respectively. Dashed salmon and dodger blue lines mark the separation of different phases of the beginner and the intermediate, respectively.

### 3.6 Joint moments

Joint moments of the beginner and the intermediate skiers are depicted in Fig. 8. Peak values for hip flexion and extension moments in this study were 1.69 Nm/kg and 2.71 Nm/kg, respectively. These values stay within physiological limits reported by Grabiner et al. [32] for hip flexion (14.84 Nm/kg) and extension (3.85 Nm/kg).

The peak knee extension moment of the beginner, executing a skidded turn, was 1.13 Nm/kg, which was lower than 2.74 Nm/kg observed for the intermediate skier performing an approximate carved turn. Interestingly, Klous et al. [33] reported a higher peak knee extension moment during a skidded turn (8.35 Nm/kg) compared to a carved turn (4.07 Nm/kg). While our study's findings align with the reported moment for the carved turn, the moment for the skidded turn was approximately 7 times lower. This discrepancy could be attributed to differences in turning radius (Klous et al. [33]: 10 m vs. our study: 7.7 m) and speed (Klous et al. [33]: 10.4 m/s vs. our study: 5.5 m/s).

The hip adduction, hip rotation, and knee flexion moments exhibited similar patterns between the beginner and the intermediate, revealing the principles of turning dynamics. Notably, there were spikes in these moments during the transition between turns. The timing of these spikes (shown in Fig. 8) coincides with the points of postural shift (shown in Fig. 5) and the shift in contact force between skis (shown in Fig. 7), indicating that muscles exert force to adjust posture and maintain balance against external forces.

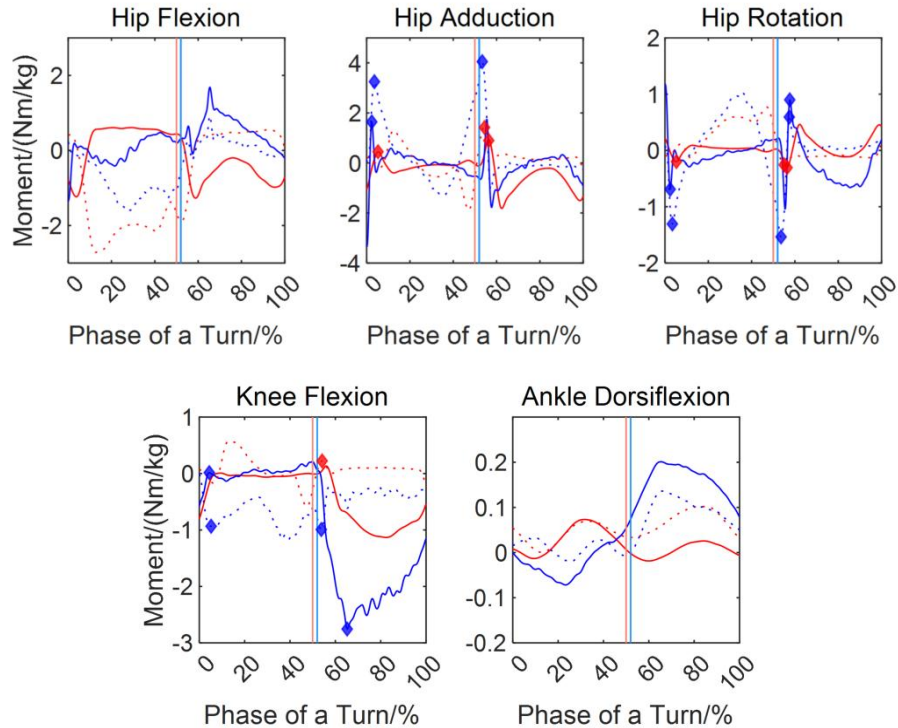

**Fig 8** Joint moments of the beginner and the intermediate. Red and blue lines indicate data of the beginner and the intermediate, respectively. Solid and dashed lines represent joint moments for the right side, and left side, respectively. Salmon and dodger blue lines indicate transition time between turns of the beginner and the intermediate, respectively. Positive moments denote: hip flexion, hip adduction, hip internal rotation, knee flexion, and ankle dorsiflexion. The diamonds represent the points where sudden spikes in joint moments occur during the transition between turns.

## 4. Discussion

Precise modeling of the ski-snow interaction is essential for an accurate analysis of a skier's force generation patterns, thereby facilitating meaningful comparisons between skiers. The accuracy of the model was validated by comparing the simulated CoM trajectory and ski-snow contact force with corresponding experimental data. The average distance error of the trajectory is 0.134 m, which is deemed acceptable given the inherent complexities and unpredictability present in real-world conditions. These factors include but are not limited to uneven terrain, varying snow conditions and fluctuating wind speed. The close alignment between our simulated trajectory and the experimental data suggests that our model effectively captures the essential physical principles and dynamics governing the skier's motion. The RMSD of the foot pressure ratio was 0.162. This disparity primarily stems from the limitations of plantar pressure insoles, which are unable to detect the force exerted by the boot cuff, focusing solely on the pressure exerted by the bottom of the ski boots. While direct comparison of magnitudes is not feasible, the force-time characteristics determined by pressure insoles and 3D-dynamometers exhibited a high similarity.[12][34] Therefore, the correlation coefficient of the foot pressure ratio was computed, which yielded a value of 0.964. This strong correlation underscores the model's capability to accurately replicate the intricate balance-shifting dynamics in skiing maneuvers, thus reinforcing its validity and utility in biomechanical analysis.

The main findings of the present study reveal the dynamics of ski turning process for both a beginner and an intermediate skier. The observed differences in velocity, which directly indicate the skill level, can be attributed to the larger steering angle of the skis and the wedge skiing posture adopted by the beginner. These kinematic variations are fundamentally linked to differences in joint moments, as motion is generated by the musculoskeletal system. This quantitative analysis offers a significant advancement over classical training methods that rely on the naked-eye observation and experience of instructors. By extracting kinematic and kinetic parameters of skiing, this study provides valuable detailed insights that can inform ski training and aid in skill improvement.

In addition to these advancements, several limitations should be acknowledged. Firstly, we did not incorporate poles into the model, thereby neglecting the twisting force introduced by pole plants. Considering that pole planting is predominantly used by advanced skiers and the participants in our study rarely employed this technique,[22] we deemed this simplification necessary due to the significant increase in model complexity associated with integrating a flexible beam model of poles. Secondly, we only validated vertical component of the ski-snow contact force due to the limitation of plantar pressure measurement insoles. This decision was made to enable participants to utilize their own equipment[12] and to ensure comfort during testing. However, for a more comprehensive verification of the ski-snow contact force, it is advisable to conduct force plate measurements to obtain 3-D forces. Thirdly, this study analyzed only two skiers performing turning maneuvers at their preferred speed and radius. To gain a more comprehensive understanding, future research should include a larger cohort of participants across various skill levels. Additionally, analyzing a broader range of skiing techniques would provide deeper insights into the dynamics of skiing.

## 5. Conclusion and outlook

This study presents an approach to capturing kinematic and kinetic data synchronously throughout the entirety of an alpine skiing run. Through the development of the SSSI model grounded in mechanical principles, we reconstructed the dynamic-interaction process in skiing. By meticulously modeling ski-snow contact behavior and incorporating factors such as gravity and air resistance, our model realistically evolved through the turns, yielding accurate skiing trajectories. Importantly, our model maintained dynamic consistency by utilizing only the skier's posture as input and without constraining translational DOFs between the pelvis and the ground.

Functioning as a digital twin of alpine skiing, our model provides valuable insights into various ski-snow contact mechanisms, including carving and skidding, enabling the analysis of skiers across different skill levels . By comparing joint angles, ski trajectories, ski-snow contact forces, and joint moments, we developed a quantitative method for evaluating ski turning techniques. On top of this, our findings may unveil shared patterns among skiers of similar skill levels, suggesting the potential for skill-level categorization based on our research. This comprehensive approach enhances our understanding of the complexities inherent in skiing dynamics and offers a robust framework for further exploration and refinement of skiing performance, while also contributing to injury prevention in skiing.